\documentclass[twocolumn,prl,superscriptaddress,floatfix,showpacs]{revtex4-1}
\usepackage[utf8]{inputenc}
\usepackage{amsmath}
\usepackage{amssymb}
\usepackage{graphicx}

\usepackage{graphics}
\usepackage{epsfig}

\newcommand{\be}{\begin{equation}} \newcommand{\ee}{\end{equation}}
\newcommand{\bea}{\begin{eqnarray}} \newcommand{\eea}{\end{eqnarray}}

\begin{document}

\title{Universality and Asymptotic Scaling in Drilling Percolation}

\author{Peter Grassberger}

\affiliation{JSC, FZ J\"ulich, D-52425 J\"ulich, Germany}

\date{\today}
\begin{abstract}
We present simulations of a 3-d percolation model studied recently by K.J. Schrenk {\it et al.} [Phys. Rev. Lett. 
{\bf 116}, 055701 (2016)], obtained with a new and more efficient algorithm. They confirm most of their results 
in spite of larger systems and higher statistics used in the present paper, but we also find indications 
that the results do not yet represent the true asymptotic behavior. The model is obtained by replacing 
the isotropic holes in ordinary Bernoulli 
percolation by randomly placed and oriented cylinders, with the constraint that the cylinders are parallel to one 
of the three coordinate axes. We also speculate on possible generalizations.
\end{abstract}
\maketitle

In spite of its mature age, the theory of percolation is still full of surprises \cite{Araujo}. A new page was 
turned open recently by Schrenk {\it et al.} \cite{Schrenk}, who revisited a model that was first studied long 
ago by Kantor \cite{Kantor}.
While Kantor had concluded that it was in the universality class of ordinary 3-d percolation, the simulations in
\cite{Schrenk} clearly suggest that it is in a different universality class. But these simulations also indicate
very large corrections to scaling. Since the simulations were done on systems with rather modest sizes and did not 
use extremely large statistics, we decided to perform larger simulations in order to check their claims. The 
result can be summarized easily: Although our estimates of critical parameters are significantly more precise than 
those of \cite{Schrenk} (and of course also of \cite{Kantor}), we fully confirm their main results. But we also 
find indications that these might not represent the true asymptotic behavior, which then would be even more
different from ordinary percolation.

The model studied in \cite{Kantor,Schrenk}, called ``drilling percolation" in the following, is very simple. Take 
a large solid block of size $L\times L\times L$ with $L\gg 1$ on a simple cubic lattice, and remove randomly 
columns of size $1\times 1\times L$, located randomly in the cube and oriented randomly but aligned with one
of the three axes. The maximal number of columns is $3\times L\times L$ (each column of fixed orientation can 
be in one of $L\times L$ positions, and there are three orientations). The control parameter is defined as 
\be
    p = \frac{\text{number of columns {\it not} taken out}}{3 L^2}.
\ee
Notice that this is not the fraction of retained sites, since one site can be in two or even three columns.
Nevertheless we expect qualitatively the same behavior as for ordinary site (``Bernoulli") percolation: 
While the non-removed parts percolate for $p=1$, they cannot percolate for $p=0$, and there must be a 
critical point $p_c$ in between.

The non-trivial aspect of the model is that it involves geometric objects of more than one non-trivial dimensionality. 
While the bulk is three dimensional, the columns (holes) have $d=1$. Thus we expect that the standard field 
theory for percolation \cite{Amit} cannot be applied without modifications. It is in this respect similar 
to models with long range correlations in the disorder \cite{Weinrib}, of which it is indeed a special and 
particularly simple case. 

The simulations were done in \cite{Schrenk} by means of two different algorithms, both of which seem however
to be less than optimal. In the present paper we shall use instead a very simple and efficient generalization
of the well known Leath algorithm \cite{Leath} for site percolation. 

The latter is a cluster growth algorithm that uses two data structures: 
(i) a bit map, where for each site $(i,j,k)$ of the lattice it is stored whether it had been tested 
($s_{ijk}=1$) or not yet ($s_{ijk} = 0$); and (ii) a queue or stack (depending on whether it is implemented 
breadth first or depth first \cite{Cormen}) that contains a list of ``growth sites", i.e. sites that had 
recently be ``wetted" (i.e., included in the cluster) and whose neighbors have to be tested whether they can 
be wetted or not. Notice that the bit map $s$ does not need to distinguish for tested sites whether the test 
had been positive (i.e. they were wetted) or not, as no site can be wetted later, if the first test was negative 
(this distinguishes site from bond percolation). Time in this growth process is discrete, if the difference
between the times when a site gets wetted and wets itself its neighbors is defined as one unit of time.

In our generalization we have to add three more arrays $X_{jk}$, $Y_{ik}$ and $Z_{ij}$ of sizes $L\times L$
each, the elements of which can assume 
three possible values. $X_{jk}=0$, e.g., means that it is not yet known whether the column parallel to the 
x-axis at position $(y=j,z=k)$ has been removed, $X_{jk}=1$ means that it has been removed, and 
$X_{jk}=2$ means that we {\it know} that it has not been removed (e.g. since some site in it has been 
wetted already). Thus at the beginning, all array elements are zero, except for the ``seed" $(0,0,0)$ where
the growth starts (implying $s_{000}=1, X_{00}=Y_{00}=Z_{00}=2$). 
Assume now a site $(i,j,k)$ is neighbor to a growth site, and is thus to be tested. If it had been tested 
already before, it has $s_{ijk}=1$ and nothing is done. Otherwise, if $s_{ijk} = 0$, we test for all three
directions whether the column passing through it is already known to be removed or not. If this is not yet known,
it is removed with probability $1-p$ (resp. kept with probability $p$), and the array element is put to 1
(resp. 2). After this, we wet the site iff $X_{jk}=Y_{ik}=Z_{ij}=2$, i.e. if and 
only if none of the three columns has been removed.

In a first set of runs we started with a point seed on lattices with $L=2^{11}$, followed the 
cluster growth breadth-first as long as the spans of all clusters in all three directions was $<L$, and recorded 
the three time-dependent observables $P(t), R(t)$, and $N(t)$. These are the probability that the cluster
grows for a time $\geq t$ (i.e., its ``chemical radius" is $\geq t$), the r.m.s. distance of growth sites
at time $\geq t$ from the seed, and the average number of growth sites (averaged over all clusters, those 
that are still growing and those that had already died). At the critical point $p=p_c$ we expect them to 
follow power laws
\be
   P(t) \sim t^{-\delta},\quad R(t) \sim t^z, \quad {\rm and} \quad N(t)\sim t^\eta
\ee
with finite-$t$ corrections, but without any finite-$L$ corrections.

\begin{figure}
\begin{centering}
\includegraphics[scale=0.30]{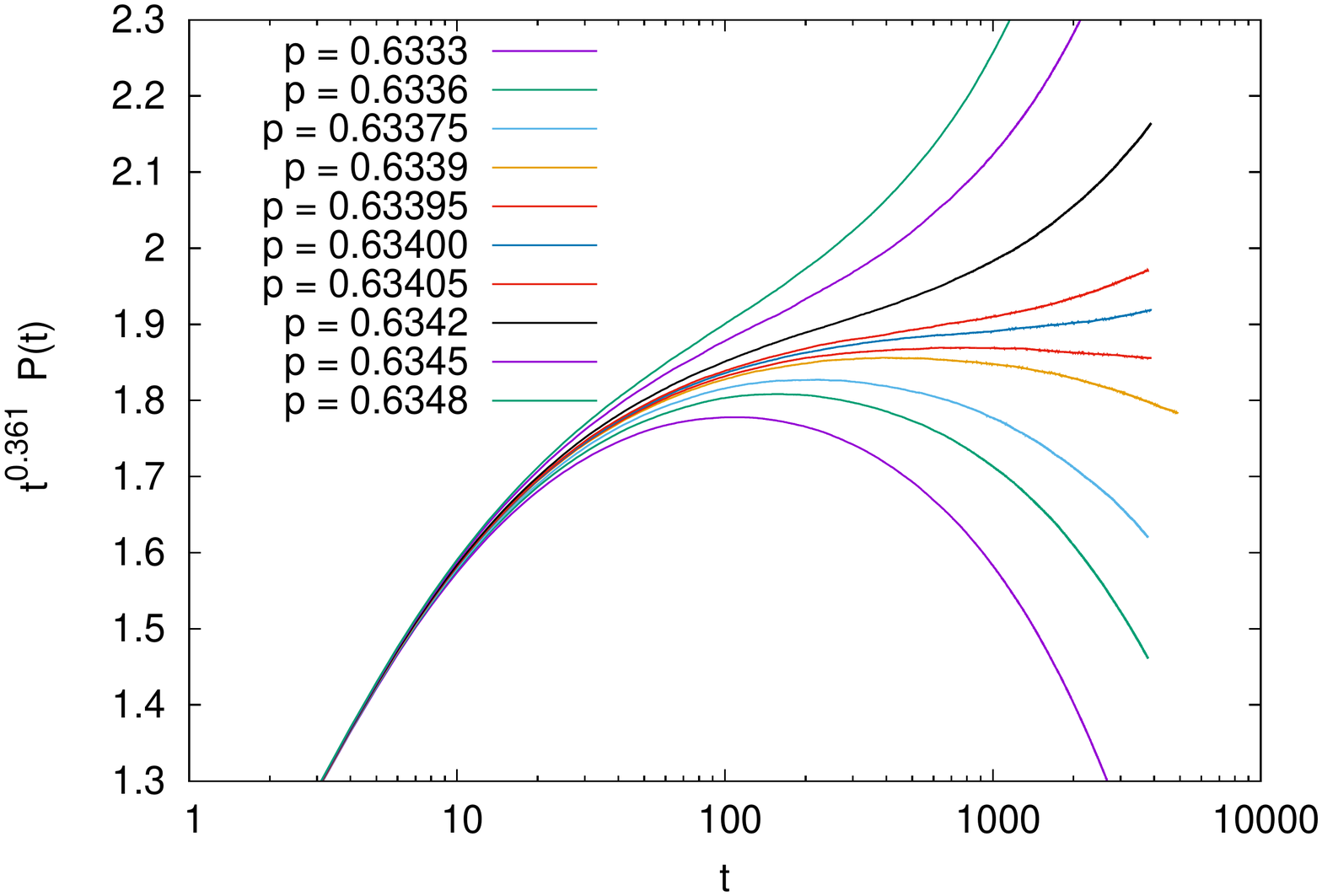}
\vglue -.9cm
\includegraphics[scale=0.30]{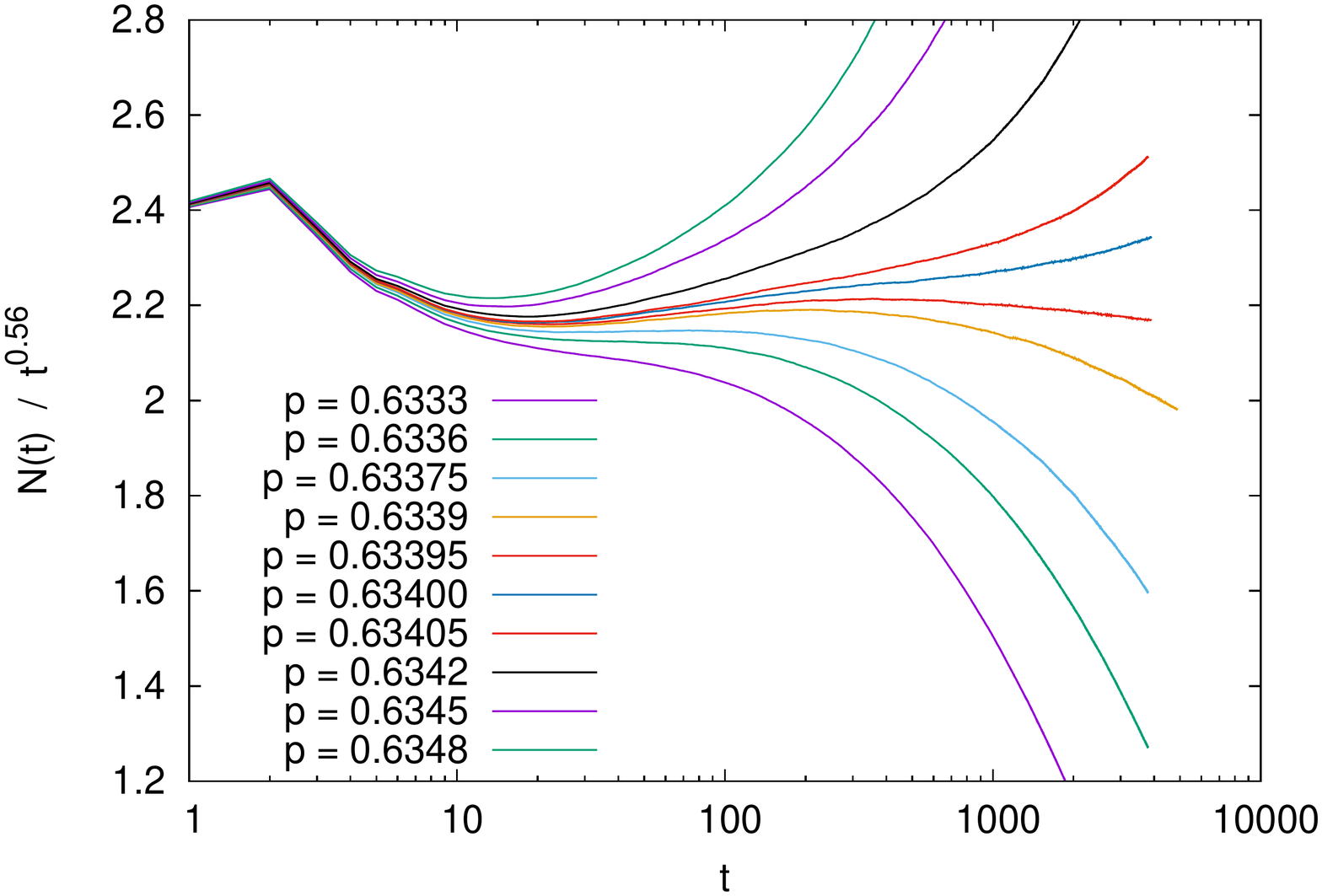}
\vglue -.9cm
\includegraphics[scale=0.30]{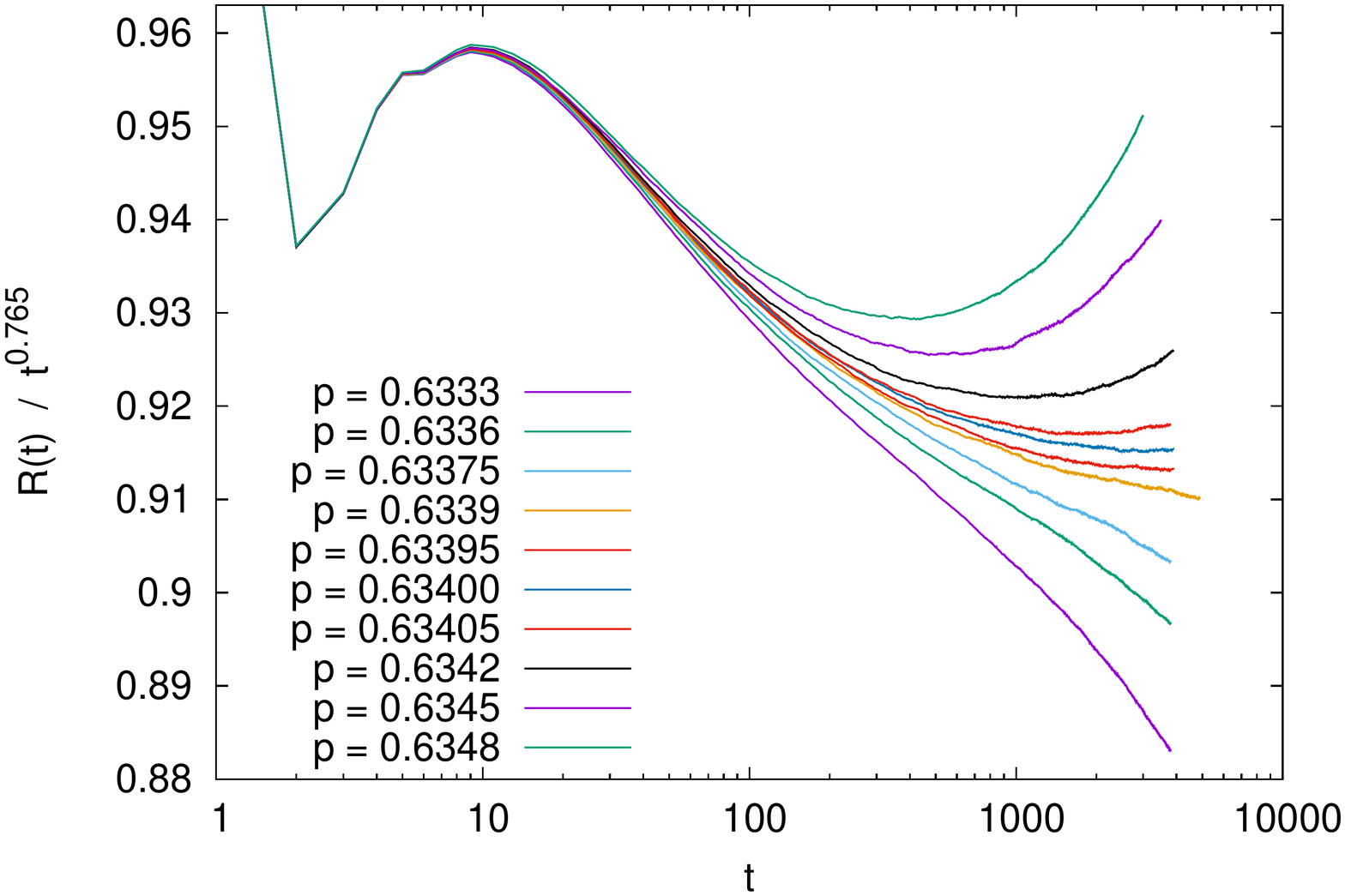}
\vglue -.8cm
\par\end{centering}
\caption{\label{fig1} (color online) Top to bottom panels: Log-log plots of survival probability $P(t)$, average 
  number of growth sites $N(t)$, and r.m.s. distance of growth sites from the seed, all plotted against time $t$. 
  Each curve is for one fixed value of $p$, with $p$ decreasing from top to bottom. For more significance, a power 
  of $t$ with suitable exponent is 
  multiplied to each curve, so that the critical curves are roughly horizontal.}
\end{figure}

In a second set of runs we used lattices with helical b.c. and followed the cluster growth until it stopped
because all wettable sites were already wetted, and measured properties like the cluster mass distribution 
$p(m)$, the dependence of the average cluster mass on $L$, and the density of the giant cluster (which is also 
the probability that a spreading from a single-site seed leads to a giant cluster).

Finally, in a third set of runs we used lattices of size $L\times L\times L_z$ with $L_z \gg L$. Initial conditions
did not consist of single ``wet" (or infected, in the interpretation of epidemic growth) points, but the entire plane 
$z=0$ was wet/infected, and the spreading was allowed
only into the region $z>0$. Lateral boundary conditions were either periodic or open, but the b.c. at $z=L_z$ was 
not specified because it was checked that all clusters stopped growing for $z < L_z$. This was feasible, because 
these simulations were only done in the subcritical phase $p < p_c$. In this way we could measure spanning 
probabilities: On a given disorder realization and for any $z_{\rm max} \leq L_z$, there exists a cluster that 
spans from $z=0$ to $z=z_{\rm max}$, iff the growth stops at $z=z_{\rm max}$. 

Results of the first (time-dependent, or ``dynamical") set of measurements are shown in Fig.~1. None of the 
curves in any of the three panels is really a straight line, indicating substantial corrections to scaling. In 
spite of this, one can identify a value $p_c \approx 0.63397$ where the curves in all three panels seem to become
straight and horizontal for large $t$. This gives us a first rough set of exponent estimates, $\delta = 0.361,
\eta = 0.56, $ and $z = 0.765$. We have not yet given error estimates, since we have two more sources of information:
The finite lattice simulations for $t\to\infty$ and, more importantly, the fan-outs of the curves in Fig.~1. 
The latter gives us an independent estimate of the correlation length exponent $\nu$. More precisely, we have 
finite-$t$ scaling laws like 
\be
   P(t,p) = t^{-\delta} \phi[(p-p_c) t^{1/\nu_t}] + \ldots,     \label{collaps_P}
\ee
and similar equations for $N$ and $R$. Here $\phi[x]$ is a scaling function that is analytic at $x=0$, and 
$\nu_t = \nu/z$.

\begin{figure}
\begin{centering}
\includegraphics[scale=0.30]{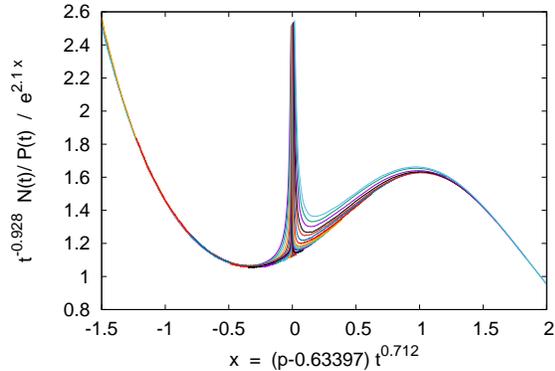}
\vglue -.8cm
\par\end{centering}
\caption{\label{fig2} (color online) Plot of $t^{-0.928} N(t)/P(t) \exp(-2.1x)$ against $x=(p-p_c)t^{1/\nu_t}$, 
  where $p_c$ is the value estimated above, and $\nu_t = 0.712$. The factor $\exp(-2.1x)$ was divided out in 
  order to reduce the range on the y-axis from about three decades to half a decade. The values of $p$ used in 
  this plot ranged from $0.623$ to $0.651$. }
\end{figure}

We checked Eq.~(\ref{collaps_P}) [and analogous ansatzes for the other observables] by plotting $t^\delta P(t,p)$
against $(p-p_c) t^{1/\nu_t}$. Parameters $p_c$ and $\nu_t$ were chosen to obtain the best data collapse. In 
view of the strong corrections to scaling seen already in Fig.~1, we cannot indeed expect a perfect collapse,
but we shall try to get a good collapse for large $t$. Results of such an attempt, this time not for $P$ but for the 
number $N/P$ of growth sites per still growing cluster, are shown in Fig.~2. This time we used a much wider 
range of $p$-values, $p \in [0.623,0.651]$. In this wide range $\phi$ would take a vast range of values, 
making a collapse plot {\it look} excellent but virtually useless. In order to increase significance, we have 
divided $\phi[x]$ by $\exp(2.1x)$. We see excellent collapse in the wings in Fig.~2, but huge deviations at 
$x\approx 0$ which precisely result from the small-$t$ corrections seen also in Fig.~1. Notice that we also
changed slightly the exponent from the above estimate, in order to optimize the data collapse.

Having obtained in this way $\nu_t\approx 1.404$, we can now also estimate other exponents like $\nu=z\nu_t$ and
$\beta= \delta \nu_t$. 

For the second set of runs we used lattice sizes ranging from $32^3$ to $2048^3$. We do not show any results,
as they were fully compatible with the dynamical simulations but proved to be less significant. In any case,
we verified that the fractal dimension $D_f$ (governing the cutoff of the mass distribution) and the exponents $\tau$ 
(describing its power law decay) are obtained in perfect agreement with the scaling relations $D_f = 3-\beta/\nu$
and $\tau = d/D_f+1$.

As a final result we obtain $p_c = 0.633965(15)$ and the mutually consistent set of exponents
$
   \delta=0.364(3), \eta=0.560(8), z=0.765(3), \nu_t = 1.404(5), \nu = 1.074(5), \beta=0.511(5), D_f = 2.524(8), 
   \tau = 2.189(5).
$
As for any critical exponent estimates, the errors here are not statistical but are mainly systematic errors due 
to uncertainties in the finite size corrections. Since critical exponent estimation involves an extrapolation 
(which by its nature is ill defined), the results are highly subjective and result from judicious plausibility
considerations taking into account all measured observables. Notice in particular that least square fits would 
not be appropriate, and are the main source of the many wrong critical exponent estimates found in the literature.
While $p_c$ is more precise than the value quoted in \cite{Schrenk} by a factor $\approx 25$, the critical exponents
are typically more precise by factors 2 to 5. But in all cases the agreement is within two standard deviations.

Let us finally discuss the spanning probabilities resulting in the subcritical phase from the third set of runs. 
Let us denote by $\Pi(p,r,L_z)$ the probability that there exists a spanning cluster (from $z=0$ to 
$z=L_z$) on a lattice with aspect ratio $r = L_z/L$, and for given $p$-value $p$.
In \cite{Schrenk} it was proven mathematically that $\Pi(p,r,L_z)$ decreases with $L_z$, for any fixed $r$ and 
for fixed $p \in [0.52974\ldots , p_c[$,
not faster than a power. This is in striking contrast to ordinary percolation, 
where $\Pi(p,r,L_z)$ decreases exponentially. A closer inspection of the proof reveals that the problem is 
similar to that of Griffiths phases \cite{Griffiths,Dhar,Moreira,Cafiero}, where frozen disorder leads to slow
decay of correlations in time. Here we are not dealing with disorder frozen in time, but with disorder (the 
columns drilled parallel to the $z$-axis) that is frozen in $z$-direction. As a consequence we should observe
correlations decreasing very slowly in the $z$-direction.

\begin{figure}
\begin{centering}
\includegraphics[scale=0.30]{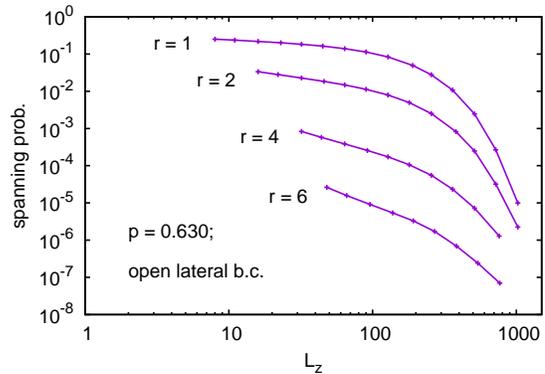}
\vglue -.8cm
\par\end{centering}
\caption{\label{fig3} (color online) Log-log plot of spanning probabilities $\Pi(p=0.630,r,L_z)$ against $L_z$,
  for fixed values of the aspect ratio $r$. }
\end{figure}

In Fig.~3 we show results for lattices with open lateral b.c. and $p=0.630$. This should be compared to Fig.~4
in \cite{Schrenk}, where the same boundary conditions and the same value of $p$ were used, but which extend 
only to much smaller values of $L_z$. Due to this much smaller range of $L_z$, the authors of \cite{Schrenk}
claimed to see a power law and thus to confirm the mathematical prediction. We now see that this was not true.
Although we cannot of course exclude an asymptotic power law, it should set in only at much larger values
of $L_z$, in particular for small values of $r$. Simulations with different values of $p$ and with periodic 
b.c.'s, not shown here, fully confirm this. 

\begin{figure}
\begin{centering}
\includegraphics[scale=0.30]{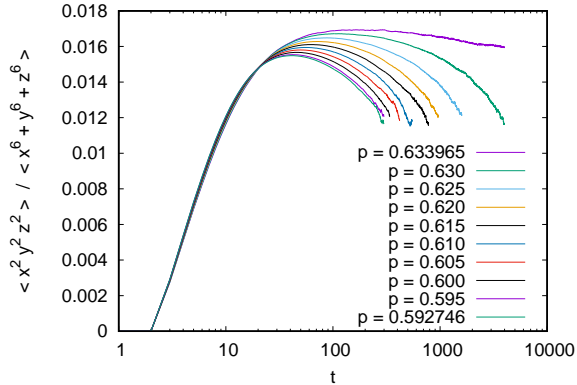}
\vglue -.8cm
\par\end{centering}
\caption{\label{fig4} (color online) Log-linear plot of a moment ratio that would be equal to $1/60$ for 
isotropic clusters. More precisely, the data use moments of the growth site coordinates at time $t$, in 
clusters grown from a single point seed. Values of $p$ decrease from top to bottom.} 
\end{figure}

Figure 3 suggests that presently reachable lattice sizes are not able to show the true asymptotic behavior.
This is also suggested by distributions of cluster sizes and of spherical asymmetries as measured by averages of
$x^2y^2z^2-(x^6+y^6+z^6)/60$, of critical and sub-critical clusters grown from point seeds. They also should 
have power-behaved tails \cite{Schrenk}. We indeed found that subcritical cluster size distributions showed 
some deviations from exponential decay (data not shown here), and that average asymmetries were significantly 
different from zero (see Fig.~4). But both were 
much smaller than what one would expect if the distributions were decaying like powers.
Thus we conclude that the above results for the critical exponents might also not yet 
represent the true asymptotic behavior -- which should be visible only for cluster and lattice sizes not 
reachable with present computational means.
The total amount of CPU time (on PC's and laptops) spent on this project was about a year.

In summary, we have verified that drilling percolation is in a new universality class, different from 
ordinary (Bernoulli) percolation, but the true asymptotic behavior might be still different from what is 
seen in \cite{Schrenk} and in the present simulations. Let us finally discuss some possible generalizations.

The most obvious generalization is mixed drilling / Bernoulli percolation, where we take out both single sites
and columns. We conjecture that in this case the long range aspect of the columns is relevant, and the 
model should be in the universality class of drilling percolation.

Next, we can consider the case where linear objects are again taken out, but they are oriented randomly, without
any reference to coordinate axes \cite{Hilario}. This seems to be more delicate. It is plausible that it is not 
in the universality class of ordinary percolation, but it might be in a universality class of its own. The same 
might be true for the case that there is a finite number $>3$ of possible orientations, e.g. coordinate axes and 
space diagonals. While the present algorithm would not work for completely random orientations, it could still 
be generalized to include diagonals.

More interesting from a theoretical point of view are higher dimensions, $d\geq 4$. For space dimension $d$ we 
can ``drill" out subspaces of dimensions $\leq d-2$ and still have a non-trivial connectedness problem. For $d=4$, 
e. g.,  taking out columns and planes would still give non-trivial percolation problems. We conjecture that these 
are in different universality classes. As one 
goes to higher and higher dimensions, one expects then a proliferation of universality classes. It seems 
however non-trivial to check this by simulations, and it is not clear how to treat them analytically.

Finally, we shall study in a forthcoming paper \cite{Grass2017} a 3-d model with columnar defects in {\it one} 
direction only and with additional point (Bernoulli) defects. This model shows a much clearer Griffiths phase, 
and much stronger anisotropies even at the critical point. 

I thank the authors of \cite{Schrenk}, in particular K.J. Schrenk, M. Hilario, and V. Sidoravicius, for 
stimulating correspondence. I also thank N. Ara\'ujo for carefully reading the manuscript.

\end{document}